\documentclass{he_symp}
\usepackage{psfig,graphicx,epsfig}
\usepackage{color}
\usepackage{amsmath,amssymb,epic,eepic,array}
\begin{document}
\renewcommand{\FirstPageOfPaper }{ 209}\renewcommand{\LastPageOfPaper }{ 214}

\title{Natural Limits for Currents in Charge Separated Pulsar Magnetospheres}
\author{A. Jessner\inst{1}, H. Lesch \inst{2} \and T. Kunzl\inst{2}}
\institute{Max--Planck--Institut f\"ur Radioastronomie,
 Radio-Observatorium Effelsberg, Max-Planck-Str. 28,  D-53902 Bad M\"unstereifel, Germany
\and Centre for Interdisciplinary Plasma Science \\
Universit\"ats-Sternwarte M\"unchen, Scheinerstr.1, D-81679 M\"unchen, Germany}
\maketitle

\begin{abstract}
Rough estimates and upper limits on current and particle densities form the basis
of most of the canonical pulsar models.
Whereas the surface of the rotating neutron star is capable of supplying sufficient
charges to provide a current that, given the polar cap potential,
could easily fuel the observed energy loss processes, observational and theoretical constraints provide strict upper limits to
the charge densities.
The space charge of  a current consisting  solely of particles having only one sign
creates a compensating potential that will make the maximum current dependent
on potential and distance. In the non-relativistic case this fact is expressed in the
familiar Child-Langmuir law. Its relativistic generalization and subsequent application
to the inner pulsar magnetosphere provides clear limits on the strength and radial
 extension of  charged currents originating on the polar cap. Violent Pierce-type
oscillations set in, if one attempts to inject more current than the space charge limit
 into a given volume. These considerations apply wherever there is a significant
amount of charged current flow, in particular in the gap regions.
There they can be used to derive limits on the size of  such gaps and their stability.

\end{abstract}

\section{Introduction}

All pulsar models require charges to be drawn from the pulsar surface, accelerated, and
possibly multiplied by pair production, as being the medium that will be responsible for
the great variety of observational features currently known from many pulsars.
Common to all models is a rotating heavily magnetized ($10^8-10^{12}$G) neutron star.
As early as 1955 it was recognized by A. Deutsch that a sufficiently strongly magnetized
fast rotator in a vacuum would be surrounded by an envelope (magnetosphere)
 of charged particles.  Because of their fundamental importance,
 the electrodynamical aspects of pulsar models have been of interest from the early days of
pulsar research until now.

In all cases, the number density of particles available is the decisive factor for the
self-consistency and hence viability of any pulsar model.

Starting from the surface of the neutron star, it is recognized that its very composition
plays a crucial role in determining the number charges emitted (Usov \& Melrose 1996).
The controlling factor is the surface work function, representing the potential barrier to be overcome by
the charges that are to be emitted from the pulsar surface.
Usually, the work function is of the order of the Fermi-energy of the solid lattice,
which in itself depends on the electron density.
Using a Thomas-Fermi-Dirac-Weizs\"acker approximation
Abrahams \& Shapiro (1992) calculated the  cold $(T=0)$
surface density $\rho_{Fe}$
to be about $2.9\cdot 10^6\, {\rm kg m^{-3}}$. That result enabled us (Jessner et. al. 2001) to estimate
the Fermi energy at the surface as

\begin{equation}
E_F(\rho_{Fe}) ={{2\cdot \pi^4\hbar^4c^2} \over {e^2B^2m_e}}
\cdot\left({{\rho_{Fe}\cdot(28-2)}\over 56\cdot m_p}\right)^2
\end{equation}
which  amounts to $E_F =4.17\cdot 10^2\, {\rm eV}$.
For such a low work function, even field emission can supply  enough charges to provide
a relativistic flow of electrons at the co-rotation value given by the Goldreich-Julian-density
(Jessner et. al. 2001).
But the estimated value for $E_F$ should be taken as an
upper limit, since the thermal X-ray emission indicates that pulsars have surface temperatures well above $3\cdot 10^5$K
and are surrounded by a thin gaseous surface layer which would
not entail any significant barrier for the extraction of electrons (Pavlov and Zavlin 1998).
In that case the surface properties themselves will not influence the primary charge densities
(free charge emission scenario).

Recently (Jessner et. al 2001), we proceeded to calculate the Lorentz factors and pair production rates of electrons accelerated
along field lines of an aligned dipole taking due account of radiation losses due to inverse Compton scattering on thermal
photons (as was proposed by Sturner 1995) and curvature radiation. This was done under the simplifying assumption of a relativistic flow which
resulted in constant shielding of the accelerating field by the primary charges below the pair production region.
In addition, the field and density enhancements described by Muslimov and Tsygan (1992)
were
also not  implemented. To be complete we will now give an estimate of the initial (primary) current that can
be drawn from a pulsar polar cap and on its stability by solving Poisson's equation with constant current density,
but with a velocity dependent charge density.

\section{The Child-Langmuir Equation}

Space charge limitations play an important role in the conduction of currents across otherwise empty space,
their detailed study in first half of the $\rm 20^{th}$ century provided the foundation of much of our electronic technology.
In order to illustrate the principle, we will briefly review a simple classical 1-dimensional case: Take two large
electrodes (Fig. 1) of equal area $A_0$ separated by a distance x, one at zero potential and the other at $U=U_0$.
\begin{figure}
\centerline{\includegraphics[width=7cm]{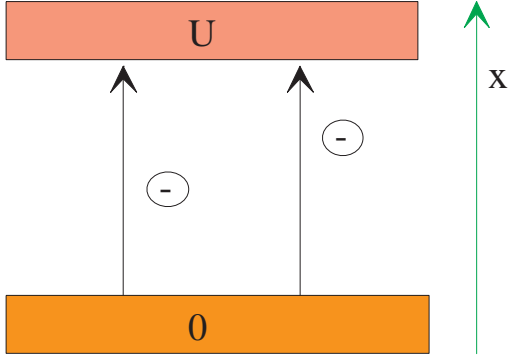}}
\caption{Simple one dimensional flow of electrons between two electrodes.
\label{image}}
\end{figure}
We assume that we can supply an arbitrary amount of charge for the flow in the direction of x,
so that the current will limit itself as soon as the potential at the lower surface vanishes.
The velocity of an electron, accelerated by a potential U is given by

\begin{equation}
\beta(U)\cdot c = { \sqrt{q_e^2 U^2 + 2q_e U m_e c^2} \over q_e U + m_e c^2 }
\end{equation}

In the non-relativistic case ($U<< 511$ kV) this simplifies to

\begin{equation}
v(U) =\sqrt{2\cdot q_e U \over m_e}
\end{equation}

If the potential $U > 0$, then a current $I$ will flow from the lower to the upper electrode.
This current is conserved along the upward (x) direction. Such a current moving in a vacuum
obeys Poisson's law:
\begin{equation}
{d^2\over dx^2}U = {I \over \epsilon_0 c A_0 \beta(U)}
\end{equation}

In the non-relativistic case we replace $c\beta$ with $\sqrt{2\cdot q_e U \over m_e}$, integrate twice
and solve for the current to obtain the
 canonical expressions for the maximum current $I_0$ and the internal resistance
$R_0=U/I_0$ of a planar vacuum diode
(Child, 1911, Langmuir 1913) that has a potential U w.r.t. the cathode.

\begin{equation}
I_0(U,x) ={4\sqrt{2} \over 3}\sqrt{q_e\over m_e} \epsilon_0 A_0 U^{3/2} x^{-2}
\end{equation}
The resistance follows as
\begin{equation}
R_0(U,x) = {3\over 4\sqrt{2} \epsilon_0 A_0 \sqrt{m_e\over q_e}} U^{-1/2} x^2
\end{equation}

Hence a very important, but often overlooked consequence of Poisson's equation and the
equation of motion is the fact,
that a charged current requires a potential to flow through a vacuum.
In the non-relativistic regime, the required potential increases as
the $4/3$ power of the longitudinal distance and the $2/3$ power of the
current.

The solution to the fully relativistic equation
\begin{equation}
{d^2\over dx^2}U = {I \cdot(q_e U + m_e c^2)  \over \epsilon_0 c A_0 \sqrt{q_e^2 U^2 + 2 q_e U m_e c^2}}
\end{equation}

is slightly more complicated,

\begin{eqnarray}
I(U,x) ={\epsilon_0 c A_0 \over q_e x^2} \Biggl( 2 \sqrt{q_e^2 U^2 + 2 q_e U m_e c^2} \\
 +m_e c^2 \left( 2 \sin^{-1}\left( {m_2 c^2 \over q_e U + m_e c^2} -\pi\right) \right)  \Biggr) \
\nonumber
\end{eqnarray}
but the general dependence of the space charge limited current on the inverse squared distance remains
unchanged. In  fig. 2 we give a quantitative example of the current that will flow between two electrodes of
$1 \rm cm^2$ crossection, spaced apart by one meter.
\begin{figure}
\centerline{\includegraphics[width=\linewidth]{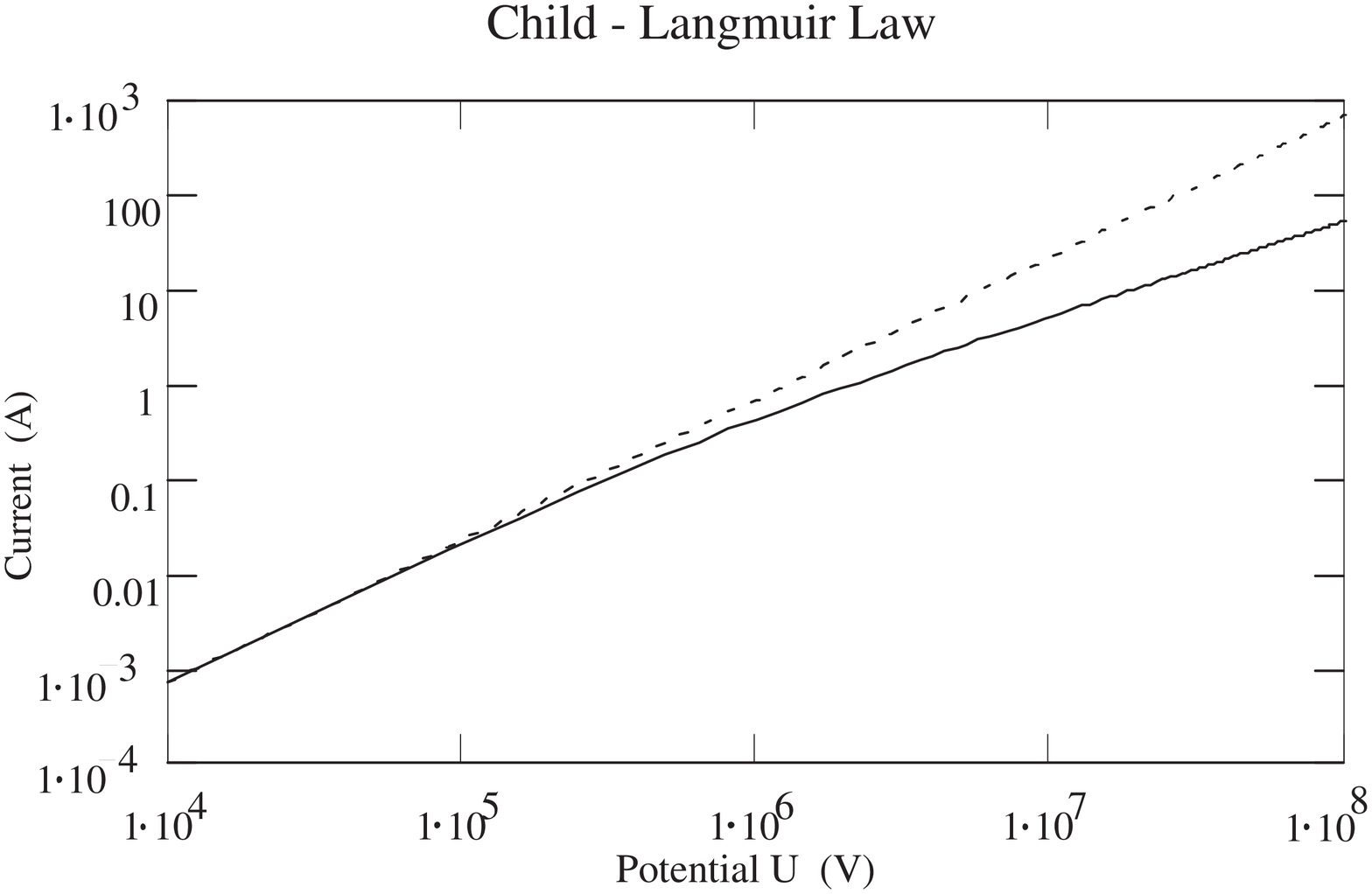}}
\caption{Current flowing between two parallel $1 \rm cm^2$ electrodes ( spacing 1 m) as a function of the
potential difference U (solid line). The dashed line shows the usual non-relativistic relationship which is invalid for
$U > {m_e c^2 \over q_e}$.
  \label{image}}
\end{figure}
In the relativistic regime, the current rises linearly with the applied potential, hence the potential required to
drive a particular current density over a distance $x$ now increases as $x^2$.
Although this example appears to be somewhat remote from the expected conditions at a pulsar's polar cap,
one easily recognizes, that for a typical Pulsar (${\rm B=10^{12}G, \ P=0.5s})$,
the relativistic Goldreich-Julian current density  approaches
\footnote{Here we used the general relativistic surface density $n(r_{ns})$  as defined in section 3}
  $q_e n(r_{ns}) c = 498 \rm\ A cm^{-2}$.
A potential of $10^9\rm\  V$ is required to drive such a current over the distance of 1 meter.
For a height of i.e. $h=100\rm m$ we have
exhausted the available potential difference ($\sim 10^{13}\rm\ V$) on the polar cap!
The square law dependence of the required potential makes it impossible to have a far reaching
{\it charged} current of appreciable density within the pulsar magnetosphere.

\section{Space charge limits for an aligned relativistic pulsar magnetosphere}
 For reasons of simplicity we will now use the model of an aligned rotator as
described by Goldreich and Julian (1969). Fig. 3 outlines the
geometry of the rotating neutron star field.
\begin{figure}
\centerline{\includegraphics[width=\linewidth]{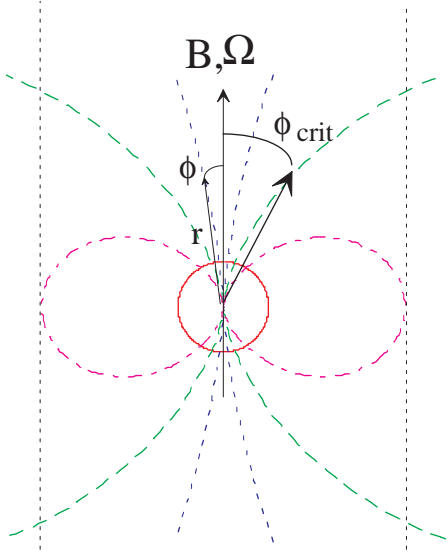}}
\caption{Aligned magnetic rotator
  \label{image}}
\end{figure}
Near the surface of the neutron star, the space-time metric is affected by the presence of the heavy compact rotator
leading to the general relativistic effects of  the Lense-Thirring dragging of field lines and a radial dependence
 of the volume elements. Muslimov and Tsygan (1992) introduced an auxiliary function to account for the general
relativistic distortion of the field lines:
\begin{eqnarray}
f(r)=-3\left({r\over r_{ns} \epsilon_{MT}}\right)^3
\biggr[
 \ln\left(1-{\epsilon_{MT}r_{ns}\over r} \right) \\
+{\epsilon_{MT}r_{ns}\over r} \left( 1+{\epsilon_{MT}r_{ns}\over2 r}  \right)
 \Biggr] \nonumber
\end{eqnarray}
here we used the compactness $\epsilon_{MT}={2\Gamma M_{ns} \over c^2 r_{ns}}$ where ${2\Gamma M_{ns}\over c^2}=r_g$ is the
canonical Schwarzschild radius of the neutron star. A magnetic field line starting with a
colateral angle $\phi_0$ at the pulsar surface is given by
\begin{equation}
\sin(\phi)=\sqrt{  {r\over r_{ns}}  {f(1)\over f\left(\ {r\over r_{ns}} \right)}  } \sin(\phi_0)
\end{equation}
 As an example we use the well known pulsar PSR B2021+51 from the catalogue of Taylor, Manchester and Lyne (1995)
 with a period of
$P=0.529\rm s$ and a surface magnetic field of $B_0=10^{12.11}\rm G$ which for simplicity we treat as if it were an
aligned rotator.
 Then the colateral angle of the rim of the polar cap turns out to be
$\phi_{crit}=\sin^{-1}\left(\sqrt{{ 2\pi r_{ns} \sin^2(\pi/2) \over P c f(1)}} \right) = 0.94^{^o}$ and the area
of the polar cap is $A_{pole}=2 \pi r_{ns}^2 (1 - \cos(\phi_{crit}))=8.5\cdot 10^4 \rm m^2$. The opening angle $\phi$ and thus
the surface area enclosed by the limiting field lines increases with increasing distance from the surface. Using the
equation of field lines (10), we can subsequenty replace $\cos(\phi(r))$ with
\begin{equation}
\zeta(r)=\sqrt{1-{r\over r_{ns}}   {f(1)\over f\left(\ {r\over r_{ns}} \right)}\sin^2(\phi_{crit})  }
\end{equation}
The Goldreich-Julian charge density compensates the induced electric field parallel to the magnetic
field line and is classically given  for the pole by
$n_0(r) = {B_0 \Omega \epsilon_0 \over q_e}\left( {r_{ns}\over r} \right)^3$
yielding $n_0(r_{ns})=8.5\cdot 10^{16}\rm m^{-3}$ on the surface of the centre of the polar cap.
 Following Beskin (1999) we can use
\begin{equation}
f_0(r)={1 -\left ({r_g\over r}\right)^3\over \sqrt{1-{r_g\over r } }}
\end{equation}
to describe  the G.R. correction of  the co-rotation density $n$ as a function of distance.
Because of the higher electric field at the surface, the  density has to increase
 by a factor of $f_0(r_{ns})=1.21$ to a value of
 $n_0(r_{ns})\cdot f_0(r_{ns})=1.03\cdot 10^{17}\rm m^{-3}$ .
$f_0(r)$ decays quickly with $r$,  i.e. to $1.021$ at $r=10r_{ns}$.
Assuming relativistic velocities, the total polar cap current can roughly account for the observed
spin-down energy loss of the pulsar:
\begin{eqnarray}
I_{cap}= n(r_{ns}) q_e c A_0 {1\over \phi_{crit}} \int_0^{\phi_{crit}} \left(3\cos^2(\theta)-1\right) d\theta\\
=1.01\cdot 10^{12}\rm A \nonumber
\end{eqnarray}
Together with the polar cap potential
$U_{cap}={-B_0\Omega r_{ns}^2 \over 2}\left( \cos^2(\phi_{crit})-1 \right) = 2.07\cdot 10^{13} \rm V$ we
obtain an electrical power of $I_{cap} U_{cap} =  2.09\cdot 10^{25} \rm W$ which, for such a coarse model,
 agrees surprisingly well with
the observed energy loss of $10^{32.9} {\rm erg\over s} = 7.9\cdot 10^{25} \rm W$.

For a pulsar with an electron current flowing out of the polar cap, the radial part of
Poisson's equation can now be written in  spherical polar coordinates:
\begin{eqnarray}
{1\over r^2}{d\over d r}r^2{d\over d r} U = {-I \over \epsilon_0 2 \pi r_{ns}^2 c \beta(u) (1-\zeta(r))} \\
+{q_e \over \epsilon_0 }n_0(r_{ns}) f_0(r) \left({r_{ns}\over r}\right)^3 (3 \zeta(r)^2 -1) \nonumber
\end{eqnarray}
The first term on the right hand side of the equation contains the charge density due to the
current of electrons and the second term the equivalent co-rotation charge density.
Here we have assumed a uniform current density across the cap and we neglect the
 transverse dependence of the
potential (all other derivatives in Poission's equation are assumed to vanish). But the divergence of field lines
as a function of distance is accounted for by $\zeta(r)$. 
Furthermore, it is quite reasonable to assume that the electrons will leave the surface with thermal velocities.
For $T=3\cdot 10^5 \rm K$ their kinetic energy corresponds to an initial potential of $U_0 =  {kT\over q_e} =25.9 \rm V$,
and a velocity of $0.01 c$. Come what may, the boundary condition at the surface $0< U < U_0$
 dictates that the flow starts in the non-relativistic regime.
As current densities are conserved along the field lines,
and charges are accelerated, the first (current) term in (14) will decrease
on acceleration, leading to further increase of the accelerating potential
(and vice versa). If we therefore have  a space charge limited flow with full shielding
at the surface ($n=n(r_{ns})$) and $\beta=0.01$,
the charge density will decrease by a factor of $\sim \beta$ along the field lines if relativistic potentials are encountered,
a point already made by Fawley, Arons and Scharleman (1977).
\section{Numerical Solutions of the Pulsar Poisson Equation}
A Bulirsch-Stoer algorithm (provided by MATHCAD) was employed with typically 2048 steps to solve (14)
for a range of up to 6m above the
pulsar surface where the boundary conditions  were set to $U(r_{ns})=U_0$ and ${d U \over dr}= 0 \rm V m^{-1}$.
The results were checked by increasing the resolution and sometimes with another algorithm
(adaptable Runge-Kutta method).
\subsection{Strict space charge limitation at the surface}
With a constant current over the polar cap that is in equilibrium at the surface, fulfilling the above boundary conditions,
we get a massive increase in particle energies with height (fig. 4).
\begin{figure}
\centerline{\includegraphics[width=\linewidth]{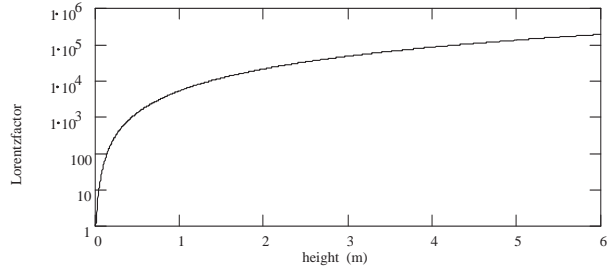}}
\caption{Lorentz factor reached by particles in a current that is limited by the space charge density being equal to
the co-rotation density at the pulsar surface.
  \label{image}}
\end{figure}
But in the same vein, the charge density falls  to about 1\% of the surface value after about 
${c \over \omega_{pl} } =c \sqrt{\epsilon_0 m_e \over q_e^2 n_0  } =15 \rm mm$ in agreement with Fawley, Arons and Scharleman (1977). 
In this and all following cases we neglected energy losses of particles by inverse Compton scattering and curvature radiation.
These losses will start to limit the attainable particle energies in the surface layer as soon as Lorentz factors of about
a few 1000 are attained (Jessner et.al. 2001).

\subsection{ Enforced current at about 0.8 times the relativistic co-rotational current }
The extremely low relativistic charge densities of $0.01 n(r_{ns})$  obtained from the steady-state solution
make pulsar models  unworkable, if  they require a  primary current of co-rotational density  (Ruderman \& Sutherland 1975)
 to supply the energy for the magnetospheric processes. Following the reasoning in section 3  we now find that
$\beta I_{cap} V_{cap} $ is more than two orders of magnitude smaller than the observed energy loss. The high, unshielded
$E_{||}$  may aid pair creation, but unless the secondary charges themselves are accelerated by an unshielded field, they
will not contribute to the energy balance.
 
It is therefore of interest to see what happens when the strict boundary
condition  $U(r_{ns})=U_0$ is relaxed by forcing a  charge overdensity
 to be drawn from the surface yielding a relativistic
current density of the order of the co-rotation density.
Fig. 5 shows the strongly fluctuating potential distribution that
will be the result.
\begin{figure}
\centerline{\includegraphics[width=\linewidth]{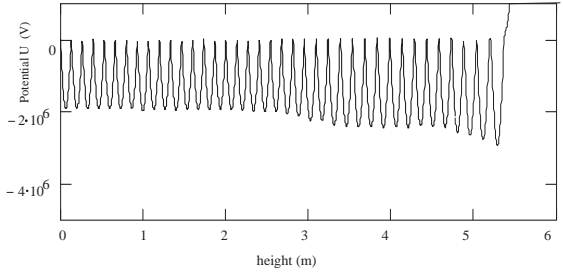}}
\caption{Effective potential along a central field line for an enforced current of about 0.8 of the relativistic co-rotation current $I_0$.
  \label{image}}
\end{figure}
The  current breaks into slabs of  $\sim \rm 10 cm$ thickness until the particles have obtained enough energy so that
their  charge density becomes low enough for the accelerating field to take over, in this case after $\sim \rm 5 m$. 
Such oscillations
are common in flows that exceed the space charge limit locally and the stationary solutions of similar cases have been described
by Mestel et. al. (Mestel et. al. 1985) and Shibata (1997).
Haeff noted the time variability of the instability in 1939, Pierce described it in 1949 and
Eilek presented (this volume) a time dependent solution showing how the oscillatory patterns evolve and propagate
with time. Schopper presented a 3-D PIC simulation (this volume) showing how an
initially steady particle beam breaks into the noted charged slabs.
\subsection{ Enforced currents with an additional  positively charged relativistic inflow  }
The local  charge overdensity on the surface, required by strong relativistic currents, can be neutralized by an inflowing current of the same
magnitude, but with opposite sign of the charges. The simplest case would be a highly relativistic current $I_r$ with
$\gamma m_e c^2 >> q_e U$. Then we can set $\beta=1$ for the return current and solve the following differential equation:
\begin{eqnarray}
{1\over r^2}{d\over d r}r^2{d\over d r} U = {-1 \over \epsilon_0 2 \pi r_{ns}^2 c(1-\zeta(r))}\left( {I\over \beta(u)} -I_r \right) \\
+{q_e \over \epsilon_0 }n_0(r_{ns}) f_0(r) \left({r_{ns}\over r}\right)^3 (3 \zeta(r)^2 -1) \nonumber
\end{eqnarray}
The positively charged current will not neutralize the space charge in an extended region above the polar cap, leading to
strong acceleration, similar to that shown in fig. 4.
But, not unexpectedly, as soon as one relaxes the neutrality condition at the surface,
density oscillations will appear. For closely matched inflowing and outflowing currents a second boundary
with $U = 0$ will exist higher up in the magnetosphere.
Fig. 6 shows an example where the effects of a positively charged  current of $0.9 I_0$ and a negative thermal current of $1.8 I_0$
are calculated.
\begin{figure}
\centerline{\includegraphics[width=\linewidth]{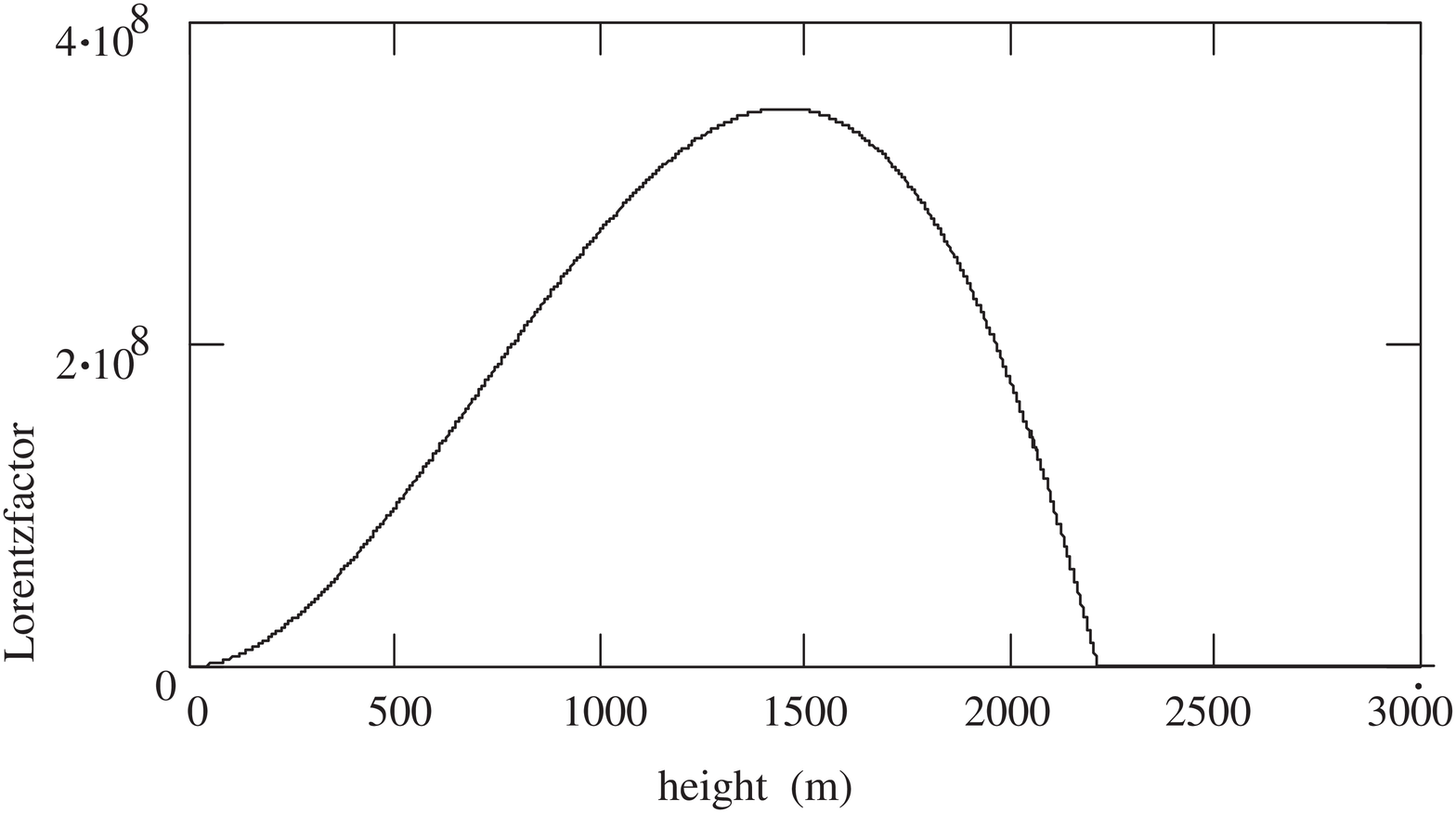}}
\caption{Lorentz factor of outflowing particles with 1.8 of the relativistic co-rotation current $I_0$ and a positive inflow
of $0.9 I_0$. Particles come to rest at $h\sim 2200\rm m$!
  \label{image}}
\end{figure}
 Further simulations with different flowrates have shown that
the location of the second boundary is very sensitive to the matching of the currents. A decrease of the outflow by one percent
moves the second boundary outward by several hundred meters.
Figs. 7 and 8 show the oscillations near the surface. Mildly relativistic potential fluctuations of a few $10^6 \rm V$
are encountered with wavelengths of about $40 \rm cm$ and the densities (Fig. 8)
\begin{figure}
\centerline{\includegraphics[width=\linewidth]{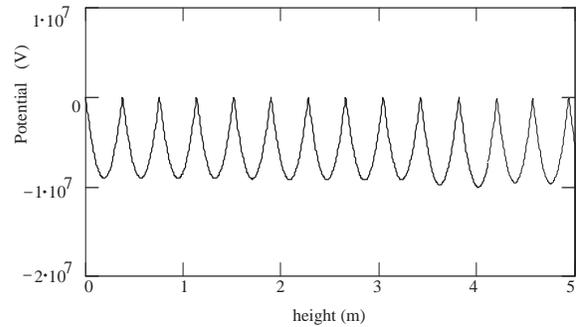}}
\caption{Potential oscillations caused by outflowing particles with 1.8 of the relativistic co-rotation current $I_0$ and a positive inflow
of $0.9 I_0$.
  \label{image}}
\end{figure}
\begin{figure}
\centerline{\includegraphics[width=\linewidth]{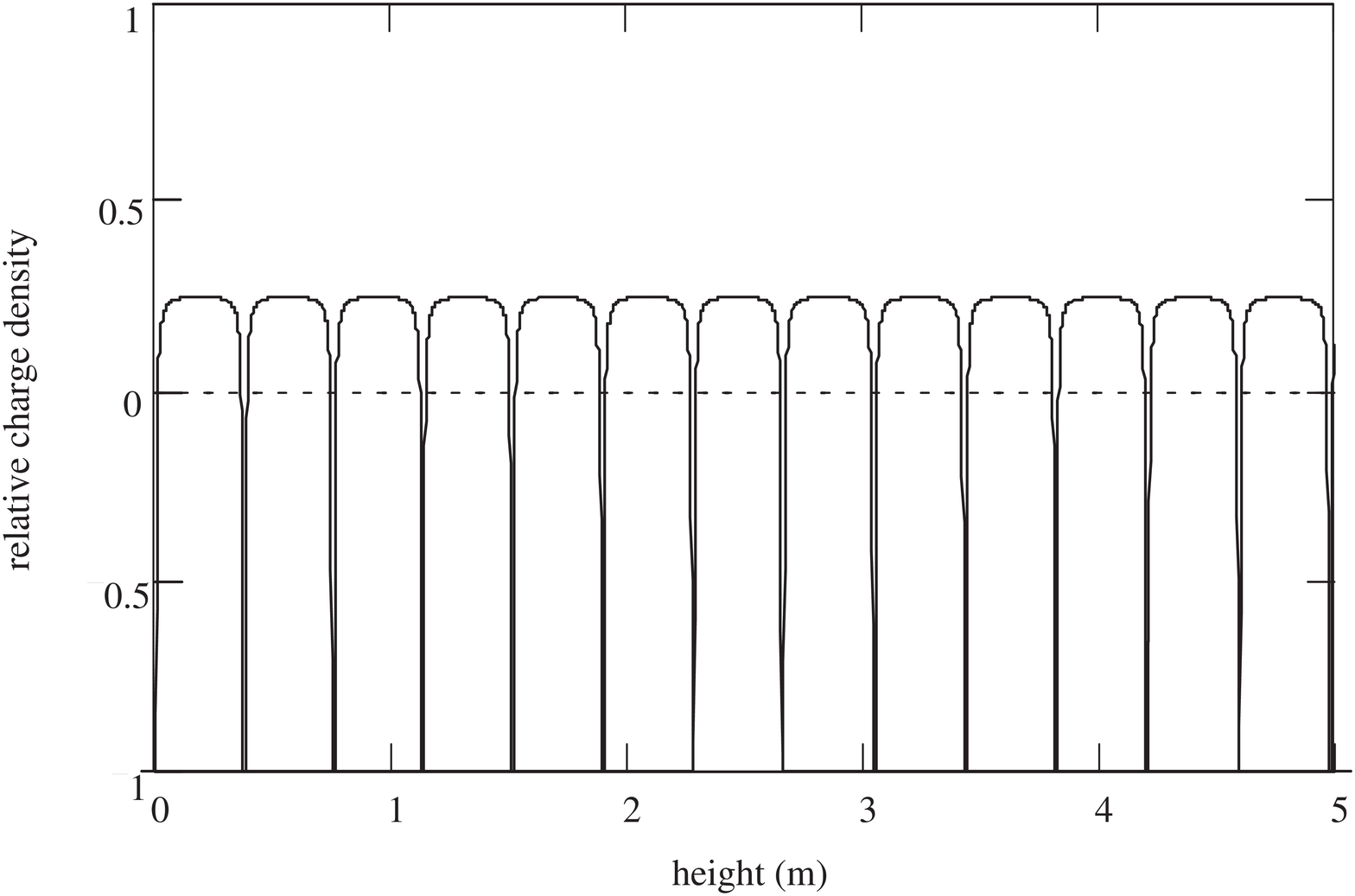}}
\caption{Charge densities for outflowing particles with 1.8 of the relativistic co-rotation current $I_0$ and a positive inflow
of $0.9 I_0$.
  \label{image}}
\end{figure}
show positive values with narrow gaps.

\section{Conclusions}
Charged currents cannot propagate in a vacuum without an outer potential, for a fixed current density the required potential
increases as the square of the distance. In the case of pulsars, the available potential on the polar cap allows for a relativistic
electron current of co-rotation density only up to a height of a few hundred metres.

Particles start from the neutron star with thermal velocities.
An electron current, if limited by non-relativistic space charges at the surface,
will encounter an outwardly increasing potential along the field lines.
This accelerates the charges and decreases the available current density. Hence a thermal current of non-relativistic surface
density cannot compensate the parallel electric field once it leaves the surface.
 In a steady-state regime,  relativistic currents from the polar cap can only be a small fraction $O(0.01)$ of a
 co-rotational current. Standard pulsar models {\it assume} a relativistic current of co-rotational density and become
unworkable when the primary currents are as low as shown. 
The available power in the magnetospheric current system is insufficient if the current is space charge limited.
Due to the unshielded $E_{||}$ Lorentz factors would become too high for
radio emission at the observed heights and if pair production were to take place, another
 boost of at least a hundred in the multiplicity factor
would be required to reach the frequently assumed $10^{3-4} n_0(r)$.
We take this as evidence against a steady state description of the pulsar mechanism and a large scale
 charged current flow out to i.e. the light cylinder.

{\it Forcing } a current that locally exceeds the  co-rotation charge
density leads to strong Pierce type oscillations. The presence of a relativistic inflow of opposite charges does not alter
the picture significantly, mainly the outflowing density can be increased by the inflowing amount.
A second boundary with $U=0$ (inner gap) can be formed for particular combinations of inflow and outflow, where the outflow
exceeds the inflow.  In these cases the
current and the potentials fluctuate with wavelengths of a few $10\rm cm$ in a region of a few metres
 above the surface. The density is strongly modulated, showing narrow gaps in the flow. Such fluctuations are inevitable
whenever the density exceeds the space charge limits.  It may happen at a boundary as in our case, or whenever a
relativistic current is strongly perturbed so that it may become non-relativistic in places and by that way exceed the
space charge limit. Work by Eilek (this volume), Schopper (also this volume) , Shibata (this volume and 1995), Mestel et. al. (1985)
does indicate that the Pierce instability is of importance  in the pulsar magnetosphere.

\vskip 0.4cm

\begin{acknowledgements}
We like to thank J. Arons, V. Beskin and A. Tsygan for drawing our attention to the fact that shielding
of $E_{||}$ cannot be uniform along the field lines. Valuable discussions with
Y. Lyubarski, R. Schopper and the particular interest in the subject shown by D. Mitra and J. Eilek were inspiring and encouraging.
We gratefully acknowledge the support by the Heraeus foundation.
\end{acknowledgements}



\clearpage

\end{document}